\documentclass[twocolumn,pra,showpacs,superscriptaddress]{revtex4}
\usepackage{graphicx,amsmath,amssymb,bbm,times}
\begin{document}
\title{Sub-shot-noise photon-number correlation in mesoscopic twin-beam of light}
\author{Maria Bondani} \email{maria.bondani@uninsubria.it}
\affiliation{National Laboratory for Ultrafast and Ultraintense
Optical Science - C.N.R.-I.N.F.M., Como, Italy}
\author{Alessia Allevi}
\affiliation{C.N.R.-I.N.F.M.-C.N.I.S.M., Dipartimento di Fisica e
Matematica, Universit\`a dell'Insubria, Como, Italy}
\author{Guido Zambra}
\affiliation{Dipartimento di Fisica e Matematica, Universit\`a
dell'Insubria, Como, Italy} \affiliation{Dipartimento di Fisica
dell'Universit\`a di Milano, Italy}
\author{Matteo G. A. Paris}
\affiliation{Dipartimento di Fisica dell'Universit\`a di Milano,
Italy}
\author{Alessandra Andreoni}\affiliation{C.N.R.-I.N.F.M.-C.N.I.S.M., Dipartimento di Fisica e
Matematica, Universit\`a dell'Insubria, Como, Italy}
\date{\today}
\begin{abstract}
We demonstrate sub-shot-noise photon-number correlations in a
(temporal) multimode mesoscopic ($\sim 10^3$ detected photons)
twin-beam produced by ps-pulsed spontaneous non-degenerate
parametric downconversion. We have separately detected the signal
and idler distributions of photons collected in twin coherence areas
and found that the variance of the photon-count difference goes
below the shot-noise limit by 3.25 dB. The number of temporal modes
contained in the twin-beam, as well as the size of the twin
coherence areas, depends on the pump intensity. Our scheme is based
on spontaneous downconversion and thus does not suffer from
limitations due to the finite gain of the parametric process.
Twin-beams are also used to demonstrate the conditional preparation
of a nonclassical (sub-Poissonian) state.
\end{abstract}
\maketitle
\section{Introduction}
Quantum correlations have been the subject of intensive
investigations and sub-shot-noise in photon-number correlations have
been indeed observed through the generation of the so-called {\em
twin-beam} of light. So far twin-beams have been obtained in the
macroscopic regime from parametric oscillators \cite{Heidman87,
Schwob97,Gao98} or from seeded parametric amplifiers
\cite{Kumar90,Raymer92} and, with much lower photon numbers, from
traveling-wave optical parametric amplifiers (OPA)
\cite{kumarTomo,JSpatial}. Quantum correlations between light beams play a
crucial role in fundamental quantum optics \cite{OL} and find applications in
quantum cryptography \cite{RFunk} and communication \cite{OptLas,Vlad},
spectroscopy \cite{Schwob97}, interferometry \cite{interf}, as well as in
enhancing sensitivity in imaging \cite{ima} and high-precision measurements
\cite{highM}.
\par
Ideal twin-beam states, as those generated by traveling-wave OPA starting from
vacuum, exhibit perfect quantum correlations in the photon number for any
average photon number, whereas approximations to them can be obtained by optical
parametric oscillators (OPO), where the increased intensity is obtained at the
price of introducing additional noise. On the other hand, existence of quantum
correlations in mesoscopic/macroscopic twin-beam states may be demonstrated
for continuous-wave OPOs via indirect measurements of the difference
photocurrent, upon spectral filtering, by analysis in a narrow frequency
bandwidth. These procedures directly give the amount of noise reduction in the
system but lose information on the number of photons in each party of the
twin-beam. Notice that OPO-based systems are unavoidably spatially multimode
\cite{spat}, as it is shown by the Poissonian distribution of the
intensities of signal and idler.
\par
The shot-noise limit (SNL) in any photodetection process is defined as the
lowest level of noise that can be obtained by using semiclassical states of
light, that is Glauber coherent states. If one measures the photon numbers of
two beams and evaluates their difference, the SNL is the lower limit of the
noise that can be reached when the beams are classically correlated. On the
other hand, when intensity correlations below SNL are observed, we have a
genuine quantum effect.
\par
In this paper we present the demonstration of sub-shot noise correlations
in a twin-beam obtained from a pulsed OPA starting from the vacuum state.
Our scheme involves a traveling-wave OPA not operated at frequency
degeneracy. The resulting twin-beam are mesoscopic (more than $\sim 10^3$
detected photons) and sub-shot noise  correlations have been demonstrated
by direct measurement of the number of photons in the two parties of the twin state.
This procedure yields complete information on the photon statistics and makes
the system available for applications, such as the production of conditional
states, which is actually feasible as described in the following. Notice that in
our system we selected single coherence twin areas in the signal and idler
light in order to match the single-mode theoretical description and
demonstrated that this experimental condition minimizes the amount of detected
spurious light and maximizes the quantum noise reduction.
\par
The scheme is conceptually simple: the signal
and idler beams at the output of the amplifier are individually
measured by direct detection. The resulting photon-counts,
$m_\mathrm{s}$ and $m_\mathrm{i}$, which are highly correlated, are
subtracted from each other to demonstrate quantum noise reduction in
the difference $d=m_\mathrm{s}-m_\mathrm{i}$. We calculate the
variance of the difference, $\sigma^2_d$ , and show that
$$\sigma^2_d < \langle m_\mathrm{s} \rangle + \langle m_\mathrm{i} \rangle,$$ where
$\langle m_\mathrm{s}\rangle$ and $\langle m_\mathrm{i} \rangle$ are
the average numbers of detected photons in the two output beams. The
quantity $\langle m_\mathrm{s} \rangle + \langle m_\mathrm{i}
\rangle$ is the SNL, {\em i.e} the value of $\sigma^2_d$ that would
be obtained for uncorrelated coherent beams.
\begin{figure}[h]
\includegraphics[width=75mm]{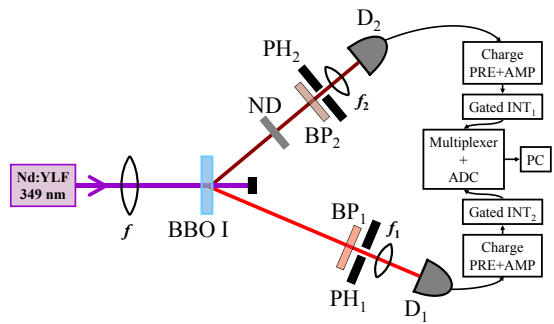}
\caption{Schematic diagram of the experimental setup. Nd:YLF,
frequency-tripled ps-pulsed laser source; BBO I, nonlinear crystal; $f$,
$f_{1,2}$: converging lenses; PH$_{1,2}$, 100 $\mu$m diameter pin-holes;
D$_{1,2}$, pin detectors; BP$_{1,2}$, band-pass filters; ND, variable
neutral-density filter. The boxes on the right side of the figure indicate the
parts of the signal amplification and acquisition chains.}\label{setup}
\end{figure}
\par
\section{Experimentals}
The experimental setup is sketched in Fig.~\ref{setup}. A
frequency-tripled continuous-wave (cw) mode-locked Nd:YLF laser
regeneratively amplified at a repetition rate of 500 Hz (High Q
Laser Production, Hohenems, Austria) produces linearly polarized
pump pulses at 349 nm with pulse duration of 4.45 ps, as calculated
from that of the fundamental pulse \cite{Paleari2004}. After passing
through a $f=250$-mm focal-length lens the pump beam comes to a focus
(diameter $\sim 300$ $\mu$m) at the position where a 4-mm-thick
uncoated $\beta$-BaB$_2$O$_4$ crystal (BBO I) is placed (Fujian
Castech Crystals, Fuzhou, China). The crystal is cut for type I
interaction at a 38.4-deg angle and it is adjusted to yield
efficient parametric amplification of a cw He-Ne laser beam at 632.8
nm that hits the crystal at a  5.85-deg angle (external angle) to
the pump beam. After the BBO I, at distances d$_1 = 60$ cm and d$_2
= 49$ cm respectively, two pin-holes of 100 $\mu$m diameter are
positioned so as to be centered with the amplified signal beam at
632.8 nm and with the idler beam at 778.2 nm. The two different
values chosen for the distances compensate the difference in the
divergence of signal and idler related to their wavelengths. Stray
light is blocked by two suitable band-pass filters, BP$_1$ and
BP$_2$, placed before the pin-holes while the light transmitted by
the pin-holes is conveyed to the detectors, D$_1$ and D$_2$, by two
25-mm focal-length lenses that are placed just beyond the pin-holes.
The interaction seeded by the cw light at the signal wavelength is
only used for the preliminary alignment of the pin-holes. The D$_1$
and D$_2$ detectors are pin photodiodes (S5973-02 and S3883,
Hamamatsu Photonics K.K., Japan) with nominal quantum efficiencies
of 86$\%$ and 90$\%$ at the signal and idler wavelengths,
respectively. As the combinations of detectors and band-pass filters
yield detection efficiencies of 55.0$\%$  in the signal arm and of
57.6$\%$ in the idler arm, the latter is lowered by adding an
adjustable neutral-density filter, ND in Fig.~\ref{setup}, in order
to obtain the same overall detection efficiency, $\eta = 0.55$, on
both arms.
\par
The current outputs of the detectors are amplified by means of
low-noise charge-sensitive pre-amplifiers (CR-110, Cremat,
Watertown, MA) followed by amplifiers (CR-200-4 $\mu$s, Cremat,
Watertown, MA), which are identical in the two arms. The two
amplified outputs are integrated over gates of 15 $\mu$s duration
synchronized with the pump pulse by Gated INT$_{1,2}$ in
Fig.~\ref{setup} (SR250, Stanford Research Systems, Palo Alto, CA).
The voltage outputs are sampled, digitized and recorded by a
computer. Accurate calibrations of the voltage outputs per electron
in the detector current pulses give 33.087 $\mu$V/el for the D$_1$ arm and
24.803 $\mu$V/el for the D$_2$ arm. All the data presented in this Letter
are in units of electrons in the D$_1$ and D$_2$ output current
pulses, thus corresponding to the number of photons detected in the
signal and idler pulses.
\par
The size of the pin-holes and their distance from the crystal have been
chosen so as to collect twin spatial modes in the downconversion
pattern. In fact, the size of coherence areas in the downconverted
light depends on the pump intensity \cite{Allevi2006,Teich96}. In
single-photon experiments this fact usually does not affect the
measurement of correlations since the probability of accidental
coincidences is very low.  On the other hand, when many photons are
involved, one should carefully select twin modes, {\em i.e.} twin
coherence areas. In our measurements, we have optimized the collection
of the light by fixing distances and sizes of the pin-holes and
maximizing correlations against the pump intensity.
\section{Sub-shot-noise correlations}
The distributions of the detected photons collected by the pin-holes are
temporally multimode \cite{Paleari2004}.
\begin{figure}[tbp]
\includegraphics[width=75mm]{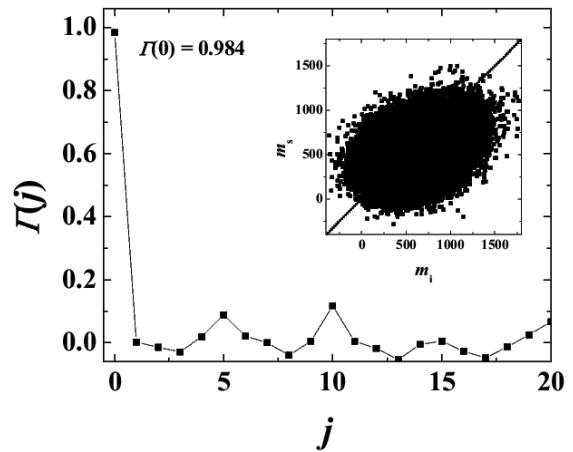}
\caption{Correlation function $\Gamma(j)$ as a function of the delay
in the laser shots (index $j$ numbers the shots). Inset: values of
the photons detected in the signal as a function of those detected
in the idler for the same laser shot.}\label{corr}
\end{figure}
In order to demonstrate that we selected twin coherence areas we
plot in the Inset of Fig.~\ref{corr} the detected photons of the
signal ($m_{\mathrm{s}}$) as a function of those of the idler
($m_{\mathrm{i}}$) for $K=10^5$ subsequent laser shots. At the pump
intensity used in this measurement, the average number of detected
photons were: $\langle m_{\mathrm{s}}\rangle=528$ and $\langle
m_{\mathrm{i}}\rangle=593$. To emphasize the presence of
correlations in the parties of the twin-beam we evaluated the
correlation function
\begin{eqnarray}
\Gamma(j)=\frac{\sum_{k=1}^K\left[\left(m_{\mathrm{s}}(k)-\langle
m_{\mathrm{s}}\rangle\right)\left(m_{\mathrm{i}}(k+j)-\langle
m_{\mathrm{i}}\rangle \right)\right]/K}
{\sqrt{\sigma^2(m_{\mathrm{s}}) \sigma^2(m_{\mathrm{i}})}} ,
\label{correl}
\end{eqnarray}
where $j$ and $k$ index the shots and $\sigma^2(m) = \langle
m^2\rangle -\langle m\rangle^2$ is the variance. Note that
Eq.~(\ref{correl}) must be corrected by taking into account the
presence of noise, both in terms of possible correlation and in
terms of increased variance. In fact, typical values of the noise
r.m.s. were $\sigma(m_{\mathrm{s},\mathrm{dark}})\sim 159$ and
$\sigma(m_{\mathrm{i},\mathrm{dark}})\sim 214$. For the data in the
inset, we obtained a value of the correlation coefficient
$\Gamma(0)=0.984$ that indicates a high degree of the correlation.
\begin{figure}[tbp]
\includegraphics[width=80mm]{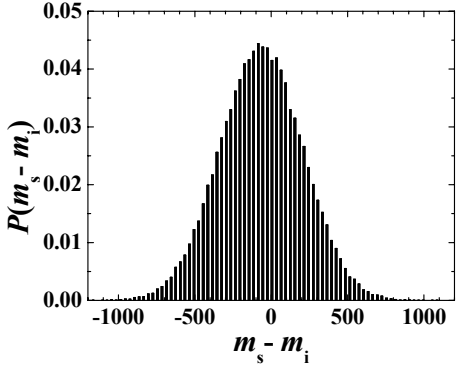}
\caption{Distribution of the difference photon-counts, $P(d)$,
measured over $10^5$ subsequent laser shots. }\label{diff}
\end{figure}
As we emphasized in \cite{Agliati2005}, a high value of correlation
is not sufficient to discriminate between quantum and classical
correlations since in both cases $\Gamma(0) \rightarrow 1$.
\par
An explicit marker of nonclassicality can be obtained by considering
the distribution of the photon-count difference $P(d)$, which is
plotted in Fig.~\ref{diff} for the same data displayed in the Inset
of Fig~\ref{corr}. The distribution appears symmetrical and centered
at zero, which indicates both accurate balance of the overall
efficiencies of the detectors and high correlation in signal/idler
photon numbers. In order to assess the nonclassicality of the
twin-beam, we calculate the quantum noise reduction $R$
\begin{eqnarray}
R=\frac{\sigma^2_d} {\langle m_{\mathrm{s}} +
m_{\mathrm{i}}\rangle}\, , \label{shotnoise}
\end{eqnarray}
where, as in the case of the correlation function, the variance of
the photon-count difference  must be corrected for the electronic
noise in the absence of light $ \sigma^2_d\rightarrow \sigma^2_d -
\sigma^2_{d,\mathrm{dark}}$. \par In principle, we have quantum
noise reduction whenever the noise reduction falls in the range
$0<R<1$ and the whole range is achievable. On the other hand, in
order to assess quantitatively the quantum noise reduction, the
overall detection efficiency $\eta$, which somewhat degrades the
observed correlations, should be taken into account. By assuming
that both arms of the measurement have the same efficiency and that
dark counts have been already subtracted, the probability
operator-valued measure (POVM) of each detector is given by a
Bernoullian convolution of the ideal number operator spectral
measure
\begin{eqnarray}
\hat \Pi_{m_j} = \eta_j^{m_j} \sum_{n_j={m_j}}^\infty
(1-\eta_j)^{n_j-m_{j}} \left( \begin{array}{c}  n_j\\ m_j
\end{array}\right)\:|n_j\rangle\langle n_j | \label{povm1}\;,
\end{eqnarray}
with $j=i,s$. The joint distribution of detected photons
$P(m_i,m_s)$ in the idler and signal beams
can be evaluated by tracing over the density matrix of
the two modes, while the moments of the distribution are evaluated
by means of the operators
\begin{align}
\widehat{m_j^p} &= \sum_{m_j} m_j^p \: \hat \Pi_{m_j}
\\
&= \sum_{n_j=0}^\infty (1-\eta)^n \:G_{\eta_j}(n_j)\:
|n_j\rangle\langle n_j |
\nonumber
\end{align}
where $G_\eta(n)= \sum_{m=0}^{n} \left(
\begin{array}{c}  n\\ m \end{array}\right)\:
\left(\frac{\eta}{1-\eta}\right)^{m}\!\! m^p$. Of course, since
$\widehat{m_j^p}$ are operatorial moments of a POVM, we have, in
general, $\widehat{m_j^p} \neq \hat m_j^p$. The first two moments
correspond to the operators
\begin{align}
\hat m_j & =\eta _j\hat n_j \\
\widehat{m_j^2} &=\eta^2_j\hat n_j^2 + \eta_j (1-\eta_j) \hat n_j
\end{align}
As a consequence, the variances of the two photocurrents are larger
than the corresponding photon number variances. We have
$$\sigma^2(m_j) = \eta_j^2 \sigma^2(n_j) + \eta_j (1-\eta_j) \langle
\hat n_j\rangle\,.$$
Using this relations we may evaluate the expected
noise reduction, which, for our multimode twin-beam obtained by
spontaneous down-conversion, is given by
\begin{eqnarray}
R=1-\eta\:,
\end{eqnarray}
independently of the gain of the amplifier.
\par
If $1> R> (1-\eta)$ the field displays nonclassical correlations
\cite{Agliati2005}, whereas $R=1$ marks the boundary between
classical and nonclassical behaviors. For the data in
Fig.~\ref{corr}, which correspond to maximum noise reduction, we get
a value $R =- 3.25$ dB below SNL. Notice that given the overall
detection efficiency $\eta=55\%$ ($1-\eta = - 3.47$ dB), this is
almost the maximum detectable noise reduction and corresponds to an
ideal noise reduction (corrected for the quantum efficiency, {\em
i.e.} measured by $100\%$-efficient detectors at the output of the
crystal) equal to $R = -14.4$ dB.  Notice also that the use of
spontaneous (not seeded) downconversion allows us to have a quantum
noise reduction that is independent of the gain of the parametric
amplifier. In fact, for an amplifier seeded by a coherent signal
\cite{Raymer92,Kumar90} the noise reduction is expected to be
\begin{eqnarray}
R=1-\eta \left( 1+ \frac{1}{2 |\nu|^2}
\frac{|\alpha|^2}{1+|\alpha|^2} \right)^{-1}\, , \label{Rseed}
\end{eqnarray}
where $\alpha$ is the amplitude of the coherent seed and $|\nu|^2$
is the gain of the OPA. For $\alpha \gg 1$ we have $$R\simeq 1- \eta
+ \eta/(2|\nu|^2)$$ and sub-shot-noise correlations may be observed
only for high-gain amplifiers.
\par
\begin{figure}[h]
\begin{center}
\includegraphics[width=80mm]{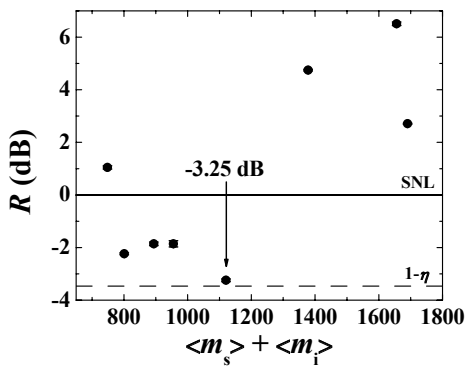}
\end{center}
\caption{Quantum noise reduction $R$ (in units of dB) as a function
of the average number of detected photons in the signal and idler
beams. Solid and dashed lines denote the shot-noise level and the
minimum value $R=1-\eta$ of the noise reduction, respectively. Error
bars are within the size of the plotted points.}\label{subshot}
\end{figure}

The study of $R$ as a function of the pump intensity represents an
useful criterion for the selection of a single coherence area. To
this aim, in Fig.~\ref{subshot} we show the values of $R$ measured
by keeping fixed the collection areas (same pin-holes located at the
same distances as for the data in the previous figures) and by
varying the intensity of the pumping beam. The figure shows that
there is an optimum condition at which $R$ is minimum. In fact,
increasing the pump intensity leads to the enlargement of the
coherence areas so that they are only partially transmitted by the
pin-holes (values on the right side of Fig.~\ref{subshot} with
respect to the minimum). On the contrary, lowering the pump
intensity reduces the size of the coherence areas and hence
introduces uncorrelated light in the pin-holes (values on the left
side of Fig.~\ref{subshot} with respect to the minimum). Note that
the values of $R$ corresponding to the selection of more than a
single coherence area remain below the SNL or quite close to it as
the information contained in the twin areas is not lost, but only
made more noisy. On the other hand, selecting only part of the areas
causes a loss of information thus determining an abrupt increase of
$R$ above the SNL. We interpret this result as an indication of the
need of perfect matching of the pin-hole areas in order to obtain sub-shot noise correlations the coherence areas with . Note that, as
the detected twin-beam is dichromatic, errors in the positioning of
the pin-holes do not affect the two parties of the twinbeam in the
same way.
\par
\par
The results reported so far demonstrate the sub-shot-noise, {\em
i.e.} quantum, nature of the intensity correlations in our
twin-beam. These features should not be confused with the presence
of entanglement, which, though expected in our system, cannot be
claimed on the basis of the present measurements (a direct evidence
of entanglement would be achievable by measuring two conjugated
quadratures by homodyne detection).
\section{Conditional measurements}
\begin{figure}[h]
\begin{center}
\includegraphics[width=80mm]{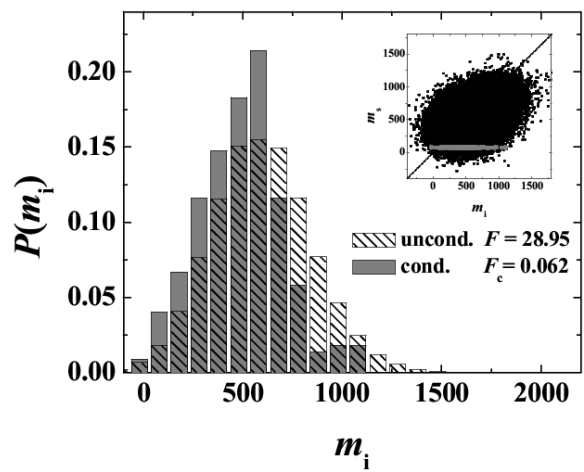}
\end{center}
\caption{Conditional photon distribution $P(m_\mathrm{i}|m_\mathrm{s}\in \Delta)$
for the idler beam (solid columns), as obtained by selecting detected
signal photon in the interval $\Delta$ indicated in the inset. We
also report the idler unconditional marginal distribution (patterned
gray).}\label{f:condM}
\end{figure}

The correlated states produced by our system are
suitable to generate mesoscopic single-beam nonclassical states by
conditional measurements \cite{CondMe,Grangier}.
The conditional measurement is made as follows: the number of
detected photons in the idler, $m_\mathrm{i}$, is kept only if the
number of detected photons in the signal, $m_\mathrm{s}$, falls into
a specific interval $\Delta$ far from its mean value. The conditional
distribution
$$P(m_\mathrm{i}|m_\mathrm{s}\in \Delta)=
\sum_{m_\mathrm{s}\in\Delta} P(m_\mathrm{i},m_\mathrm{s})$$
is reported in Fig.~\ref{f:condM} together with the
unconditional marginal distribution $P(m_\mathrm{i})=\sum_{m_\mathrm{s}} P(m_\mathrm{i},m_\mathrm{s})$,
$P(m_\mathrm{i},m_\mathrm{s})$ being the joint idler-signal photon distribution.
The narrowing of the conditional distribution below the SNL is
apparent. The Fano factor ($F = \sigma^2(m)/\langle m\rangle$) of
the idler conditional
distribution, suitably corrected for the noise, is given by
$F_{\mathrm{c}}=0.062$ (to be compared to that of the corresponding
unconditional marginal distribution $F=28.95$), which corresponds to
a noise reduction of about $12$ dB below the single-beam shot-noise
level. The probability of success, {\em i.e.} the fraction of data
that are kept to build the distribution is $0.22\%$. The success
probability may be increased by enlarging the interval of
acceptance, at the price of increasing the Fano factor, which goes
above unity if the acceptance window is too large or closer to the
mean value of the unconditional distribution \cite{CondMe}.
\par
\par
Our detection scheme, being based on direct detection of the photons
in the two arms of the twin-beam, is somehow more noisy as compared to the schemes adopted in \cite{Raymer92, Kumar90}, which make use of electronic
subtraction of the pin photocurrents of signal and idler. On the
other hand, our technique has the advantage of keeping the
information about the marginal probability distributions, whose data
can be used for state conditioning/engineering by post-selection on
one of the two parties. Of course, when the information on the
marginal distributions is not required, the direct measurement of
the difference photocurrent is preferable. Our technique, being
based on the ability of properly selecting the twin coherence areas,
provides a system suitable for many applications exploiting the
sub-shot-noise correlations of the twin-beam
\cite{lane,rarity,shelby,fabre,apb01}. Indeed, the twin-beam may be
used as a maximally effective probe to reveal either tiny
displacements imposed by light-matter interactions or by external
perturbations. For instance, one can implement absorption
measurements of weakly absorbing samples by looking at the
degradation of the noise reduction or at the variation of the Fano
factor in conditional measurements, thus improving the performances
of single-mode techniques \cite{Nap}. In this framework, the fact
that the twin-beam is dichromatic, and hence its two parties are
tunable in frequency, is undoubtedly relevant.
\section{Conclusions and outlooks}
In conclusion, we have demonstrated the possibility of obtaining
sub-shot-noise intensity correlations in the ps-pulsed mesoscopic
regime by properly selecting the twin coherence areas on the parties
of the twin-beam. Our system, which is based on spontaneous
downconversion, does not suffer from limitations due to the finite
gain of the parametric process and allows us to demonstrate
conditional generation of sub-Poissonian light. The mesoscopic
nature of our twin-beams makes them a promising system to reveal the
presence of small perturbations, {\em e.g.} weakly absorbing
biological samples, in a non-destructive illumination regime
suitable to preserve even photolabile compounds.
\par\noindent
This work has been supported by MIUR projects PRIN-2005024254-002
and FIRB-RBAU014CLC-002.

\end{document}